\documentclass[prl,twocolumn,groupedaddress,noshowpacs,nofootinbib]{revtex4}
\usepackage{graphicx}
\usepackage{epsfig}

\def\C{{\cal C}}
\begin{document}
\title{Majorana Neutrino Superfluidity and Stability of Neutrino Dark Energy}

\author{Jitesh R. Bhatt}
\email{jeet@prl.res.in}

\author{Utpal Sarkar}

\affiliation{Physical Research Laboratory, Ahmedabad
380009, India}

\begin{abstract}\
We demonstrate that Majorana neutrinos can form Cooper pairs due
to long-range attractive forces and show BCS superfluidity in a
class of mass varying neutrino dark energy models. We describe the
condensates for Majorana neutrinos and estimate the value of the
gap, critical temperature and Pippard coherence length for a
simple neutrino dark energy model. In the strong coupling regime
bosonic degree of freedom can become important and Bose-Einstein
condensate may govern the dynamics for the mass varying neutrino
models. Formation of the condensates can significantly alter the
instability scenario in the mass varying neutrino models.
\end{abstract}

\maketitle

Some time back neutrino superfluidity was studied for Dirac
neutrinos \cite{kap04}, in which the left-handed and the right-handed
neutrinos form Cooper pairs due to the attractive force originating
from their Yukawa interactions with the Higgs scalar. 
Unfortunately this interesting concept could not be applied to
any realistic situation in astrophysics or cosmology.
We extend this formalism for Majorana neutrinos and show
that the superfluidity of relic neutrinos could be important
when one considers their interactions with very light scalar
field, like quintessence. 
We study the superfluidity of Majorana neutrinos 
in the context of the mass varying neutrinos (MaVaNs) models 
\cite{mavans,mavans1} and show that it can solve the stability
problem \cite{stab1,stab2} of the MaVaNs {\sl naturally}.

The interaction between the neutrinos
in a MaVaN scenario is known to be attractive\cite{de1,hill,hill1}
due to the presence of a quintessence field called acceleron.
It is well known that these models are unstable when the neutrinos
become non-relativistic {\it i.e.} their pressure $p_{\nu}\approx 0$
 \cite{stab1,stab2}. The instability saturates
when the degeneracy pressure balances the attractive
force and the final state can evolve as $\Lambda$CDM. 
It should be noted that
this instability does not arise in a certain class of models
involving super-acceleration \cite{raman} 

We show that this attractive interaction 
can lead to neutrino superfluidity in MaVaNs by formation of 
the Cooper pairs.
If the size of the Cooper pairs is smaller than the
length scales relevent for the dark energy dynamics,
the dynamics of a scalar field describing the
Bose-Einstein condensate could be applied for studying the evolution of 
the system. 
The inclusion of the condensate dynamics alters the instability scenario
significantly. Firstly, there would be  no degeneracy pressure
in the bosonic system. Moreover the coupling between the neutrinos
and the scalar field will be changed. However the attractive
force, if any, between the condensates and the acceleron  
should be balanced by the Heisenberg uncertainty.
This kind of stable structures are known in the literature as Boson stars
\cite{kaup}. Next we demonstrate that the 
stability calculations considered earlier\cite{stab1,stab2} are altered
and the new stability criteria can be satisfied 
by the different models of dark energy potentials.
Condensates with Majorana neutrinos has been
discussed in the literature \cite{ju}, but here we develope 
a statistical formalism following ref. \cite{kap04}.

The first problem one encounters while dealing with the 
Majorana neutrino supefluidity is the chemical potential.
Since the Majorana particles self annihilate, number of particles 
is not conserved, and hence, their number operator and the
chemical potential vanishes in equilibrium.
However, in the early universe, these are not of any concern:
the Majorana neutrinos have two helicity components which
can vary with time according to the helicity-flip rate.
One can define the chemical potential to the extent the
helicity is conserved. Typically the ratio of
the helicity-flip rate to the current Hubble expansion $H$
rates is 
\begin{equation}
\frac{G^2\,T^3gm^2_\nu}{\sqrt{g}T^2/m_p}\sim 10^{-8},
\end{equation}
where $g$ is the effective degrees of freedom relevent for the
 temperature $T$ \cite{lang}.
Thus the helicity-flipping rate of the Majorana particles ceases below a certain
temperature, when the particles move apart from each other
due to the expansion of the universe at a faster rate compared
to their self annihilation rate \cite{majochem}.

Another important problem is to show how the freely streaming streaming neutrinos
can become degenerate and exhibit superfluidity:
In any MaVaN model there is an interaction between the scalar
field and the neutrinos. Dynamics of the neutrinos can be described by the
following kinetic equation[Afshordi et.al. in Ref.\cite{stab1}] 
\begin{equation}
 \frac{df}{d\eta}+{\bf u}\cdot{\bf \nabla}f-a\gamma^{-1}{\bf \nabla}\,m_\nu\cdot
\frac{\partial\,f}{\partial\,{\bf p}}=0
\end{equation}
\noindent
where $a$ and $\eta$ are the scale factor and the conformal time, defined in Robertson-Walker 
metric, respectively. The last term includes the effective neutrino mass variation due to the scalar field. In the absence of the last term on the left hand side, $f$ is
given by the usual Fermi-Dirac distribution. However, when 
small perturbations of the type 
$\delta f = \Delta(p)exp[i({\bf k}\cdot{\bf x}-\omega\eta]$ and 
$\delta m_\nu=\Sigma exp[i({\bf k}\cdot{\bf x}-\omega\eta]$ 
are considered, the system becomes unstable and eq.(1) gives, in a sub-Hubble regime,
\begin{equation}
\omega \Delta({\bf p})={\bf k}\cdot{\bf u}-\gamma^{-1}\left({\bf k}\cdot\frac{\partial f}{\partial 
{\bf p}}\right)\Sigma 
\label{dispafs}
\end{equation}
Only $\omega/k=c_s$ appear in the equation and there is no preferred scale in this equation. $c_s$ can be found (Afshordi et. al. in Ref.\cite{stab1}) to be $\pm\sqrt{-1}$ in a nonrelativisitic
regime. Thus the instability that grows at smaller length 
scale. Since there is no preferred length scale in equation (\ref{dispafs}), the instability can continue to grow on the smaller length scales. However in a realistic situation the instability may saturate when the degeneracy pressure by the neutrinos become important\cite{we1}. At this stage the attractive interaction between  the
degenerate neutrinos may induce the phenomenon of superfluidity.

Consider a Majorana mass term of the left-handed neutrinos in a model that 
provides an attractive long-range force due to the exchange of a
scalar field $\phi$:
\begin{equation} \label{a1}
{\cal L}_M = \overline{\nu_M} ~[m_\nu + f_\phi ~\phi] ~\nu_M \,.
\end{equation}
%
Although we present
the formalism for one gneration, it can be easily extended
to the realistic case with three neutrinos.


The Majorana field $\nu_M$, defined in terms of the left-handed
neutrinos $\nu_L$ and its CP conjugate field ${\nu^c}_R$:
\begin{equation}
    \nu_M = \nu_L + \lambda {\nu^c}_R
\end{equation}
satisfies the condition:
\begin{equation}
 \nu_M^c = \left[\nu_L + \lambda {\nu^c}_R \right]^c = \lambda^\ast \nu_M.
\end{equation}
where $\lambda$ is the Majorana phase, $|\lambda|^2 =1$. We work in
the Weyl representation:
\begin{eqnarray}
    \gamma^\mu &=& \pmatrix{0 & \sigma^\mu \cr \bar \sigma^\mu & 0}
    \hskip .15in {\rm with} ~~ \sigma^\mu = [I_2,\sigma_i]; ~
    \bar \sigma^\mu = [I_2,-\sigma_i] \nonumber
\end{eqnarray}
where $I_2$ is a $2 \times 2$ unit matrix and $\sigma^i$ are the
Pauli matrices. In this basis, $\gamma_5 = i \gamma^\circ \gamma^1
\gamma^2 \gamma^3 = {\rm diag}\pmatrix{-I_2, & I_2}$ is diagonal, 
and the left and right-handed fields become:
\begin{eqnarray}
    \nu_L = \pmatrix{ \psi \cr 0};
    \nu_R = \pmatrix{ 0 \cr \overline{ \chi}};
    {\nu^c}_R = \pmatrix{ 0 \cr \overline{ \psi} };
    {\nu^c}_L = \pmatrix{ \chi \cr 0}.
\end{eqnarray}
The Majorana neutrinos can now be expressed as:
\begin{eqnarray}
    \nu_M = \pmatrix{
    \psi \cr \lambda \bar \psi};~~
    \nu_M^c = \pmatrix{
    \lambda^\ast \psi \cr \bar \psi};~~
    \overline{\nu_M}^T = \pmatrix{
    \lambda^\ast \overline{\psi}^\dagger \cr \psi^\dagger};
\end{eqnarray}
so that the Lagrangian density associated
with the mass term (equation \ref{a1}) becomes
\begin{eqnarray}
    {\cal L}_M &=& [m_\nu + f_\phi \phi] \overline{\nu_M} ~\nu_M
    = [m_\nu + 
f_\phi \phi] \left[\lambda^\ast \overline{\psi}^\dagger \psi
    + \lambda \psi^\dagger \overline{\psi} \right]. \nonumber
\end{eqnarray}
It ought to be mentioned that the present mass of scalar field
differ widely from $10^{-4}eV$ as in Fardon {\it et. al.} in
Ref.\cite{mavans} 
to $m_\phi >\,H$ {\it e.g} Bjaelde {\it et. al.} in Ref. \cite{stab2}. 
In what follows, we deviate from the initial model of Fardon {\it et. al.} and consider
the case with $m_\phi > H$.
For small energy and momentum transfers, interaction term can be
written as
\begin{equation}
    H_I = - \C ~(\overline{\nu_M} ~\nu_M) ~(\overline{\nu_M} ~\nu_M)\,.
\label{int1}
\end{equation}
In terms of the component fields $\psi$ this becomes
\begin{eqnarray}
    H_I &=& - \C  \left[
    {\lambda^\ast}^2 ~\overline{\psi}^\dagger_{a} ~\psi_{a}
    ~\overline{\psi}^\dagger_{b} ~\psi_{b}
    + \overline{\psi}^\dagger_{a} ~\psi_{a} ~{\psi_{b}^\dagger} ~\overline{\psi}_{b}
    \right. \nonumber \\&+&
    {\psi^\dagger_a} ~\overline{\psi}_a~ \overline{\psi}^\dagger_{b} ~\psi_{b}
    + \lambda^2 ~{\psi^\dagger_a} ~\overline{\psi}_a~
    {\psi^\dagger_{b}} ~\overline{\psi}_{b}
    \left.   \right] \,.
\label{int2}
\end{eqnarray}

One of the key ingredient in theory of superconductivity is
to have an overall attractive interaction between its particles.
In the case of a metal  the Coulomb interaction between the electrons,
in Fourier space is $V_{Coul}=e^2/(k^2+K_D^2)$, where ${K_D}^{-1}$
is the typical shielding distance. $V_{Coul}$ is always 
repulsive {\it i.e.} $V_{coul}>0$. However the superconductivity arises as
the the electrons in the metal
also have an attractive interaction $V_{ph}$ arising 
due to their interaction with the phonons. One can show that
for a superconductor $V_{ph}+V_{Coul}<0$. Under this condition 
it is energetically more favourable for particles to form pairs.
Momenta of the particles in the pair states are directed opposite
with each other and have spin in the opposite directions.
When the energy minimization is carried out for the wave function
containing the pair states either occupied by the two particles
or none the gap condition naturally arises \cite{fey}.
In the case of MaVaN scenario, as implied by Eqs.(\ref{int1}-\ref{int2}) there is
an attractive interaction between the neutrons. 
In fact the famous instability in this scenario arises
In fact the instability
in MaVaN scenario arises precisely when this interaction dominates
over the gravity \cite{stab1}. Since there is no other interaction
that can make up for the overall repulsive interaction,
the above condition of superconductivity can be satisfied. 

The gauge boson exchange would give repulsive force between two
left-handed fields, so the only possible condensate would correspond
to a spin-0 pairing of the left-handed neutrinos with the
right-handed antineutrinos:
\begin{equation}\label{a2}
    \langle \psi_{a} ~\overline{\psi}^\dagger_{b} \rangle = \epsilon_{ab} ~D \,.
\end{equation}
The mean field approximation would then give us the interaction
Hamiltonian with the condensate $D$:
\begin{equation}
    H_1^{MF} = -2 ~\C \left[ {\lambda^\ast}^2 ~
    \overline{\psi}^\dagger_{a} ~{\psi}_{b}~ D + \lambda^2~
    {\psi^\dagger_a} ~\overline{\psi}_{b}~D^\ast \right]~\epsilon_{ab} \,.
\end{equation}
We shall now express the Majorana field in terms of the
creation and the annihilation operators as
\begin{equation}
\psi_M (x) = \sum_{p,s} \sqrt{m_\nu \over 2 \epsilon } \left(
f_{ps} u_{ps} e^{-i p x} + \lambda^* f_{ps}^\dagger v_{ps} e^{i p x}
\right) \,.
\end{equation}
The component fields are then related to the creation and
annihilation operators through the relation
\begin{eqnarray}
    \psi =  \sum_{p,s} \sqrt{m_\nu \over 2 \epsilon }
    f_{ps} u_{ps} e^{-ipx} && \overline{\psi}
    =  \sum_{p,s} \sqrt{m_\nu \over 2 \epsilon }
    f_{ps}^\dagger v_{ps} e^{ipx} \nonumber \\
    {\psi}^\dagger =  \sum_{p,s} \sqrt{m_\nu \over 2 \epsilon }
    f_{ps}^\dagger \bar u_{ps} e^{ipx}
    && \overline{\psi}^\dagger
    =  \sum_{p,s} \sqrt{m_\nu \over 2 \epsilon }
    f_{ps} \bar v_{ps} e^{-ipx} \,. \nonumber 
\end{eqnarray}
The interaction Hamiltonian can then be written in terms of
the creation and annihilation operators as:
\begin{eqnarray}
    H_1^{MF} &=& -\C  \sum_{p} {m_\nu \over \epsilon} \left[
    D ~ {\lambda^\ast}^2~ e^{-2i \epsilon t} \left(
    f^{\phantom{\dagger}}_{p \uparrow}
    f_{-p \downarrow} - f_{p \downarrow}
    f_{-p \uparrow} \right)
    \right. \nonumber \\
    &&+ D^\ast ~ {\lambda}^2~ e^{2i \epsilon t} \left.
    \left( f^\dagger_{p \uparrow}
    f^\dagger_{-p \downarrow} - f^\dagger_{p \downarrow}
    f^\dagger_{-p \uparrow} \right)
    \right] \,,
\end{eqnarray}
where $\epsilon = \sqrt{p^2 + m_\nu^2}$.

In models of mass varying
neutrinos the number density of the Majorana neutrinos becomes
proportional to the inverse of neutrino mass. In addition, the
effective interaction Hamiltonian also does not conserve particle
number. 
Since the treatment is based on grand canonical ensemble,
this requires a self-consistent treatment to determine
when the condensates become nonvanishing.
The complete Hamiltonian (${ H}$) 
is obtained by adding  the interaction part $H_1^{MF}$
and the free-particle Hamiltonian
\begin{equation}
    H_0 = \sum_{p} \epsilon~ \left( f^\dagger_{p \uparrow}
    f_{p \uparrow} + f^\dagger_{p \downarrow}
    f_{p \downarrow} \right) \,.
\end{equation}
One can also write the complete Hamiltonian in a so called 
standard or canonical form in which it resembles
with the free particle Hamiltonian in Eq.(15) \cite{super} as 
\begin{equation}
    {\cal H} = \sum_{p} E~ \left( b^\dagger_{p \uparrow}
    b_{p \uparrow} + b^\dagger_{p \downarrow}
    b_{p \downarrow} \right) \,,
\end{equation}
where $E^2 = (\epsilon - \mu)^2 + \kappa^2$ and
$\mu$ is the chemical potential.
 
For the late universe when the neutrino become 
non-relativistic, $T\ll m$,
one can write its chemical potential following  Ref.\cite{Kolb}
as 
$$
 \mu(t)=m\,+\,(\mu_D-m)T(t)/T_D \nonumber$$
where $T_D$ is the decoupling temperature and  $T(t)$ can be
regarded as the current temperature. In the late universe the
second term on the right hand side can be negligible compared
to the first term.

One can have a time-dependent transformation that can relate
the complete Hamiltonian $H = H_0 + H_1^{MF}$ with the standard
form given by Eq.(16) \cite{kap04,super}. A relation between the annihilation and
the creation operators in both the Hamiltonians is given by
\begin{eqnarray}
  b_{p \uparrow} &=& \cos \theta e^{i(\alpha + \epsilon t)}
  f_{p \uparrow} - \sin \theta e^{i(\alpha + \epsilon t)}
  f^\dagger_{-p \downarrow} \nonumber \\
  b_{p \downarrow} &=& \cos \theta e^{i(\alpha + \epsilon t)}
  f_{p \downarrow} + \sin \theta e^{i(\alpha + \epsilon t)}
  f^\dagger_{-p \uparrow}\,,
\end{eqnarray}
\noindent
with $D {\lambda^\ast}^2 = |D | e^{2i \alpha}$,
$\tan 2 \theta = \kappa/(\epsilon - \mu)$  and
$\kappa = 2~\C |D |m_\nu/\epsilon$.
A consistent solution
for the nonvanishing condensate $D \neq 0$ requires $\alpha = \pi/2$,
which has contributions from both the condensate $D$ as well as from
the Majorana phase $\lambda^\ast$. The magnitude of the gap is
determined by the consistency condition that the value of the condensate
is same as that of the value obtained by the canonical transformation.
In other words, if we express the condensate in terms of the density
matrix ($\rho$) as
\begin{equation}
    \langle \psi_{a} ~\overline{\psi}^\dagger_{b} \rangle =
    \rho \psi_{a} ~\overline{\psi}^\dagger_{b} \,,
\end{equation}
the density matrix ($\rho$) satisfies the consistency condition
\begin{equation}
    \rho = {e^{-\beta H - \mu N} \over \sum e^{-\beta H - \mu N}}
    = {e^{-\beta {\cal H} } \over \sum e^{-\beta {\cal H} }} \,.
\end{equation}
This condition translates into
\begin{equation}
    {\C \over 2} \int {d^3 p \over (2 \pi)^3} {m_\nu^2 \over \epsilon^2} {1 \over
    \sqrt{(\epsilon - \mu)^2 + \kappa^2}} = 1 \,,
\end{equation}
\noindent
whose solution gives us the magnitude of the gap.
This integral is divergent and it should be cut off with
the upper limit $\Lambda$.
The condensates form due to the attractive force between
the neutrino and the scalar field in the MaVaN scenario.
For the early times when the neutrinos were in thermal contact
with the other species in the universe, this attractive interaction
may not be very important. Thus the values 
of $\Lambda$ 
can be estimated from the energy scales below which the attractive 
interaction can be felt by the neutrinos become important. 

Solving this equation we obtain the gap
\begin{equation}
    \Delta = 2 \sqrt{2 \Lambda \over m_\nu} \left(
    3 \pi^2 n_\nu \right)^{1/3} e^{-x} \,,
\end{equation}
where $ x = 2 \pi^2 / [\C m^2_\nu (3 \pi^2 n_\nu )^{1/3}]$. The
critical temperature and the Pippard coherent length are given
by
\begin{eqnarray}
  T_c = {e^\gamma \over \pi}\Delta \approx 0.57 \Delta;  &~~~~&
  \xi = {e^x \over \pi \sqrt{2 \Lambda m_\nu}}\,.
\end{eqnarray}
This completes the formalism of formation of Cooper pairs with
Majorana neutrinos and BCS superconductivity. This may have
many applications.

We shall now discuss how the Majorana neutrino superfluidity,
can solve the stability problem \cite{stab1,stab2} in neutrino
dark energy models \cite{mavans,mavans1,hill}.
We demonstrate this in a specific two-generation neutrino 
dark energy model \cite{hill}. 
The standard model is extended with two right-handed neutrinos
$N_i, (i=1,2)$ and two scalars $\Phi_i, (i=1,2)$ with a global
$U(1)_1 \times U(1)_2$ symmetry, so that these fields interact
as
\begin{equation}
{\cal L}_M = {1 \over 2} \alpha_1 \bar N_1 N_1^c \Phi_1
+ \alpha_2 \bar N_2 N_2^c \Phi_2 .
\end{equation}
When the fields $\Phi_i$ acquire vacuum expectation values ($vev$)
$\langle \Phi_i \rangle = f_i$, we can express them as
\begin{equation}
    \Phi_i = {f \over 2 \sqrt{2}} e^{2 i \phi_i/f} \,,
\end{equation}
where $\phi_i$ are the massless Nambu-Goldstone bosons
and we assumed same decay constant $f_i = f$ for both the fields. 
Writing $\alpha_i \Phi_i = M_i {\rm exp}[2 i \phi/f]$,
we get the masses of the right-handed neutrinos $N_i$ to be $M_i$.

In this model, the neutrino Dirac mass terms
$m_{ij} \bar \nu_i N_j$
do not respect the global symmetry, and hence, one combination of
the global $U(1)$ symmetries is broken explicitly. As a result,
one of the two massless Nambu-Goldstone bosons picks up a small
mass, making it a pseudo Nambu-Goldstone boson (pNGB). This pNGB
(denoted by $\phi$) can then become the acceleron field that
explains the dark energy 

After integrating out the heavy right-handed neutrinos, we can
write down the mass matrix of the light physical neutrinos $\nu_p^T =
\pmatrix{\nu_1 & \nu_2}$, as given in equation \ref{a1}, 
where $f_\phi = - i m^2/2Mf$ is not diagonal, leading to a long-range
attractive force, and $m_p = m^2/M$.
In these models of MaVaNs, naturalness restricts the mass scales of the model
and the right-handed neutrino mass scale is supposed to be
as low as eV. 

To estimate the parameter  $x$ we need to know $\C$ which depends
on the acceleron mass $\C\sim\frac{1}{8m_{\phi}}$. If one takes
$m_{\phi}\sim 6 Mpc$ \cite{stab2}, $m_\nu\sim 1 eV$ and $n_\nu\sim 56/cm^3$
one finds $x\sim 10^{-56}$. Thus practically the
exponential factors in Eqs.(21-23) are unity. So, the relevent scales of
this equations are determined by the cut-off $\Lambda$ and neutrino mass
$m_\nu$. There remains a great deal of uncertainty over the range of
these parameters. For example if one can take $\Lambda$ as the scale
when the tracking $\rho_\nu\sim \rho_{DE}$ becomes valid
[see Fardon {\it et.al} in Ref.\cite{mavans}], one can
take $\Lambda$ as the decoupling temperature $1\,MeV$. One can also
take $\Lambda$ to be very close to the scale when the neutrinos 
become non-relativistic {i.e.} few times $m_\nu$. If we take the neutrino mass around $1\,eV$,
$\xi$ has range between $0.36-10^4\, cms$. The values of $\xi$
in this entire range still can be smaller than the neutrino lumps
\cite{we1}.
Finally we comment on the so called  instability in MaVaN models\cite{stab1}
in this changed scenario with the condensates.
The instability is known to arise when the
coupling between the scalar field and the neutrinos is stronger than the gravitational
force. The coupling  can be described by a source
term of the type $\beta(\phi)(\rho_{\nu}-3p_{\nu})$ \cite{stab2}. In the the relativistic 
regime $\rho_\nu\sim 3p_\nu$, the coupling is highly suppressed. However, it can
become strong in the non-relativistic limit $p_{\nu}\sim 0$.
Size of the Cooper pairs is determined by the interaction strength between the scalar
field and the neutrinos. 
$\phi_s$ dynamics
is given  by the following Lagrangian density
\begin{equation}
{\cal{L}}=\partial_{0}\phi_s^{\dagger}
\partial_{0}\phi_s\,-
\partial_i\phi_s^{\dagger}\partial_i\phi_s -V(\phi_s),
\end{equation}
\noindent
where, $m\simeq 2m_\nu$  represents mass of the condensate.
The condensate potential 
\begin{equation}
V(\phi_s)=m^2\vert\phi_s\vert^2+g\vert\phi_s\vert^4 
\label{poten}
\end{equation}
There can be interaction between the
condensate $\phi_s$ and the scalar field $\phi$ given by $V_{int}=g_1\phi\phi_s$. 
We can write down the perturbation equation for the coupled system $\phi$ and
$\phi_s$, following Bjaelde, {\it et. al.} in Ref.\cite{stab2}, for the 
minimum of the effective potential tracked by the fields,
\begin{eqnarray}
\delta\ddot{\phi}+2H\delta\dot{\phi}+\left[k^2+a^2V^{\prime\prime}_\phi\right]
\delta\phi&=&g_1\delta\phi_s \label{pert1}\\
\delta\ddot{\phi_s}+2H\delta\dot{\phi_s}+\left[k^2+a^2V^{\prime\prime}_{\phi_s}\right]
\delta\phi_s&=&g_1\delta\phi.
\label{pert2}
\end{eqnarray}
\noindent
It should be noted that $g_1$ quantifies the coupling between the condensate field
$\phi_s$ and the scalar field $\phi$ and it can be much smaller
than the coupling between the neutrinos and the scalar field.
The condensates already formed due to the attractive between the Majorana neutrino mediated by 
the $\phi$. 

Eqs.(\ref{pert1}-\ref{pert2}) contain $V^{\prime\prime}_\phi$ and $V^{\prime\prime}_{\phi_s}$ terms
which are non-linear functions of $\phi$ and $\phi_s$ respectively. These  equations
are linearized by making the assumption that they  can be written as 
$\phi\,\approx\,\phi_0+\delta\phi$ and $\phi_s\,\approx\,\phi_{s0}+\delta\phi_s$,
where, the quantities with suffix $0$ represent background quantities and they
can vary with time much slowly than the perturbation. 
From this perturbation in the scalar field 
\begin{equation}
\delta\phi= \frac{g_1\delta\phi_s}
{-\omega^2+2iH+k^2+a^2V^{\prime\prime}_\phi}
\end{equation}
\noindent
We can compute $V^{\prime\prime}_\phi$ in Eq.(24) using two forms of potential $V_\phi$
frequently used in the MaVaN models. First we consider Coleman-Weinberg
type of potential \cite{coleman}
\begin{equation}
V_\phi\,=\,V_0log(1\,+\,\kappa\phi)
\end{equation}
\noindent
where parameters $V_0$ and $\kappa$ can be selected to yield total dark energy contribution
$\Omega_{DE}\simeq 0.7$. Using the above linearization one can find $V^{\prime\prime}_\phi
=\frac{2\kappa^2V_0}{\left( 1+\kappa\phi_0\right)^2}$. Secondly we consider an inverse
power-law model of the potential given by
\begin{equation}
V_\phi\,=\,\frac{M^{n+4}}{\phi^n}
\end{equation}
where the parameter $M$ can be fixed by the requirements $\Omega_{DE}\simeq 0.7$ and $m_\phi\gg H$.
Thus we write as for $n<1$,
\begin{equation}
V^{\prime\prime}_\phi\,=\,\frac{n(n-1)(n-2)M^{n+4}}{\phi_0^{n-1}}
\end{equation}

Instability is known to occur in a MaVaN scenario for $H<k/a<m_\phi$ \cite{stab2}.
The following dispersion relation can be obtained from Eqs.(\ref{pert1}-\ref{pert2}):
\begin{equation}
\omega^4-4iH\omega^3-\left[4H^2+B\right]\omega^2+
2iHB\omega+B_1B_2-g_1^2=0
\label{disp}
\end{equation}
\noindent
where, $B=2k^2+a^2\left(V^{\prime\prime}_{\phi}+
V^{\prime\prime}_{\phi_s}\right)$, $B_1=k^2+a^2V^{\prime\prime}_{\phi}$
and $B_2=k^2+a^2V^{\prime\prime}_{\phi_s}$.
Eq.(30) is quartic in $\omega$ and it can be solved exactly by
analytical means.
However it is instructive to solve it by approximate methods.
The terms that are linear in the expansion rates $H$ will contribute
to the damping of the modes. Moreover the instability, in the
MaVaN scenario without the condensates, the instability was found to
occur in the regime $H<k/a<m_\phi$. Thus $H$ defines the smallest 
wave-vector of the perturbations. In what follows we  ignore the
terms involving $H$ in Eq.(\ref{disp}) to get following biquadratic
dispersion relation,
\begin{equation}
\omega^4-B\omega^2+B_1B_2-g_1^4\approx 0
\end{equation}
\noindent
Solution of this can be written as
\begin{equation}
 \omega^2\,=\,B\,\pm\,\sqrt{B^2-4\left(B_1B_2-g_1^2 \right)}
\end{equation}
\noindent 
None of the roots of Eq.(32) is imaginary if $B\ge\,0$ and
$B_1B_2-g_1^2\ge\,0$. These are the stability criteria
and their validity may depend on the form of the scalar field
potential as $B$ and $B_1$ both 
involve the term $V^{\prime\prime}_\phi$. If $B\ge 0$ and
$B_1B_2-g_1^2<0$ then the two roots of Eq.(39) are purely imaginary
and one of it could give the instability. It is easy to verify that
for the Coleman-Weinberg potential given by Eq.(33) $V^{\prime\prime}_\phi$
is positive for the parmeters values 
$V_0\approx\,8.6\times 10^{-13}eV^4$, 
$\kappa\approx\,1\times\,10^{20}M_{pl}^{-1}$ and $\phi_0\approx 10^{-6}M_{pl}$
taken from Ref. \cite{stab2}. For the condensate potential with $g>0$,
one can have all the stability conditions satisfied for a sufficiently
small coupling strength $g_1$. The stability condition satisfied for
the entire regime $H<k/a<m_\phi$. 
For the power-law kind of potential given by Eq.(34)
one can take $M\approx 0.011eV$, $n=0.01$ and 
$\phi_0\approx 0.001M_{pl}$ \cite{stab2},one can have $B_1>0$.
For this case the stability conditions are again satisfied for a
sufficiently small values of the coupling $g_1$ and $g>0$.

The Majorana neutrino superfluidity we discussed may have some interesting
consequences. The effects of the
attractive long force, required for the formation of condensates, have been
discussed for the
neutrino oscillation experiments \cite{long} and also in cosmology
\cite{hill}. The acceleron potential can also change some of the features of the
neutrino oscillations \cite{conseq}, which will be further modified in the presence
of the condensates.

We would like to note here that the stability analysis provided in this paper is for the scalar field $\phi_s$ 
describing neutrino condensates. This is a valid description when the Bose-Einstein condensation (BEC)
is formed. However when the coherent length is much smaller than the neutrino interparticle
spacing, the phenomenon of superfluidity of Majorana neutrino still occur in BCS as shown
in Ref.(\cite{kap04}). 
Furthermore, the accelerated expansion can make the ratio of the helicity flip rate to
the Hubble expansion rate even more smaller than the one we have at present.
Thus the condition for defining chemical potential in accelerated universe will remain 
satisfied in future. However the chemical potential thus defined is a 'dynamical quantity'.
But the question of the small variation in the chemical potential with time may not
be studied by the formalism presented here.

In summary we have discussed two issues for MaVaN models namely 
the superfluidity of Majorana neutrinos and the stability
of MaVaN dyanmics.
We proposed a formalism to have condensates with Majorana
neutrinos. 
Our formalism
shows that the neutrino superfluidity naturally arises in MaVaN scenario
and is a generic feature of interacting Fermi particles having
attractive potential between them at a low temperature \cite{laugh}.
In addition, we have also shown that for the case when the condensate 
dynamics become important and $m_\phi > H$ the dynamics of the condensate 
can be stable for a variety of the dark energy potentials.


\begin{thebibliography}{99}
\bibitem{kap04} J.I. Kapusta, Phys. Rev. Lett. {\bf 93}, 251801 (2004).

\bibitem{mavans} P. Gu, X. Wang, and X. Zhang,
Phys. Rev. D {\bf 68}, 087301 (2003); R. Fardon, A.E. Nelson, and N.
Weiner, JCAP {\bf 0410}, 005 (2004); P.Q. Hung, hep-ph/0010126;
R. D. Peccei, Phys. Rev. \textbf{D 71}, 0235727 (2005).

\bibitem{mavans1}
H. Li, Z. Dai, and X. Zhang, Phys. Rev.
D \textbf{71}, 113003 (2005); V. Barger, P. Huber, and D. Marfatia
Phys. Rev. Lett. \textbf{95}, 211802 (2005); A.W. Brookfield, C. van
de Bruck, D.F. Mota, and D. Tocchini-Valentini, Phys. Rev. Lett.
\textbf{96}, 061301 (2006); A. Ringwald and L. Schrempp, JCAP
\textbf{0610}, 012 (2006); R. Barbieri, L.J. Hall, S.J. Oliver,
and A. Strumia, Phys. Lett. B \textbf{625}, 189 (2005);
R. Takahashi and M. Tanimoto, Phys. Lett. B \textbf{633}, 675
(2006); R. Fardon, A.E. Nelson, and N. Weiner, JHEP \textbf{0603},
042 (2006);
E. Ma and U. Sarkar, Phys. Lett. B \textbf{638}, 356 (2006);
K. Ichiki and Y. Keum, JHEP, 000(2008); arXiv:0803.3142 (2008).
\noindent
\bibitem{stab1}
N. Afshordi, M. Zaldarriaga, and K. Kohri, Phys.Rev. D \textbf{72},
065024 (2005). It is possible that a new kind of instability
would arise even in a non-adiabatic regime in dynamical-dark-energy: 
J. V\"{a}liviita, E. Majerotto and R. Maartens,  JCAP {\bf 0807}, 020 (2008).
\noindent
\bibitem{stab2} O.E. Bjaelde, {\it et. al.}, arXiv:0705.2018v2[astro-ph];
C. Wetterich, Phys. Lett. \textbf{ B 655}, 201 (2007);
D.F. Mota, V. Pettorino, G. Robbers and C. Wetterich,
arXiv:0802.1515v1[astro-ph], see also
R. Bean, E.E. Flanagan and M. Trodden, New J.Phys {\bf 10},
03306 (2008).
\noindent
\bibitem{de1} C.T. Hill, D.N. Schramm, 
J.N. Fry, Nucl. Part. Phys. {\bf 19}, 25 (1989);
J.A. Frieman, C.T. Hill, R. Watkins, Phys. Rev. {\bf D 46}, 1226 (1992);
A.K. Gupta, C.T. Hill, R. Holman, E.W. Kolb, Phys. Rev. {\bf D 45}, 441 (1992);
J.~A.~Frieman, C.~T.~Hill, A.~Stebbins and I.~Waga,
Phys.\ Rev.\ Lett.\  {\bf 75}, 2077 (1995);
E. Masso, F. Rota, G. Zsembinszki, Phys. Rev. {\bf D 70}, 115009 (2004);
E. Masso, G. Zsembinszki, JCAP {\bf 0602}, 012 (2006);
P.Q. Hung, E. Masso, G. Zsembinszki, JCAP {\bf 0612}, 004 (2006),
\noindent
\bibitem{hill} C.T. Hill, I. Mocioiu, E.A. Paschos, and
U. Sarkar, Phys. Lett. B \textbf{651}, 188 (2007).
\noindent
\bibitem{hill1} P.H. Gu, H.J. He,
and U. Sarkar, Phys. Lett. B \textbf{653}, 419 (2007); JCAP {\bf 0711}, 016 (2007);
P.H. Gu, arXiv:0710.1044 [hep-ph].
\noindent
\bibitem{raman}
M. Kaplinghat and A. Rajaraman,
Phys. Rev. {\bf D 75}, 103504 (2007).
\noindent
\bibitem{de} C. Wetterich, Nucl. Phys. B \textbf{302}, 668 (1988); P.J.E.
Peebles and B. Ratra, Astrophys. J. {\bf 325}, L17 (1988).
\noindent
\bibitem{kaup} D.J. Kaup, Phys. Rev. {\bf C 172}, 172 (1968);
F.E. Schunck and E.W. Mielke, Class. Quant. Grav. {\bf 20},
R301 (2003).
\noindent
\bibitem{ju} S. Antusch, J. Kersten, M. Lindner and M. Ratz,
Nucl. Phys. {\bf B 658}, 203 (2003);
G. Barenboim, arXiv:0811.2998[hep-ph].
\bibitem{lang} P. Langacker, G. Segr\'e and S. Soni,
Phys. Rev. {\bf D 26}, 3425 (1982).
\noindent
\bibitem{majochem} M. Fukugita and T. Yanagida,
Phys. Rev. {\bf D 42}, 1285 (1990), M. Li, X. Wang, B. Feng
and X. Zhang, Phys. Rev. {\bf D 65 }, 1103511 (2002).
\noindent
\bibitem{fey}
R.P. Feynman, {\it Statistical Mechanics: A set of Lectures},
(Addison-Wesley Publ. Co., Redwood City, 1972).
\noindent
\bibitem{super}
A.A. Abrisikov, L.P. Gorkov and I.E. Dzyaloshinkii,
{\it Methods of Quantum Field theory in Statistical Physics} 
(Prentice-Hall, Englewood Cliffs, 1963),
L.E. Reichl, {\it A Modern Course in Statistical Physics} 
(John Wiley \& Sons, New York, 1998).
\noindent
\bibitem{Kolb}
E. Kolb and M. S. Turner,  {\it The Early Universe}, p.70,
(Addison-Wesely Publishing Company, 1990).
\noindent
\bibitem{we1} N. Brouzakis, N. Tetradis, and C. Wetterich, 
arXiv:0711.2226,[astro-ph],
P.P. Avelino, L.M.G. Beca and  C.J.A.P. Martins, Phys. Rev. 
{\bf D 77}, 101392 (2008).
\noindent
\bibitem{coleman}
S.R. Coleman and E. Weinberg, Phys. Rev. {\bf D 7}, 1888 (1973).
\noindent
\bibitem{long}  J.~A.~Grifols and E.~Masso,
  Phys.\ Lett.\ B {\bf 579}, 123 (2004);
V. Barger, P. Huber, and D. Marfatia, Phys. Rev. Lett.
{\bf 95}, 211802 (2005); M. Cirelli, M. C. Gonzalez-Garcia, and C.
Pena-Garay, Nucl. Phys. {\bf B719}, 219 (2005);
 M.~C.~Gonzalez-Garcia, P.~C.~de Holanda, E.~Masso and R.~Zukanovich Funchal,
  arXiv:hep-ph/0609094.
\bibitem{conseq} D.B. Kaplan, A.E. Nelson, and N. Weiner, Phys. Rev. Lett.
\textbf{93}, 091801 (2004); G. Dvali, Nature {\bf 432}, 567 (2004).
\noindent
\bibitem{laugh}
R.B. Laughlin, {\it A Different Universe}, (Basic Books, New York, 2005).
\end{thebibliography}
\end{document}